%% file: main.tex
\documentclass[10pt,aps,prx,twocolumn,superscriptaddress]{revtex4-2}

\usepackage[a4paper,margin=1.25cm]{geometry}
\usepackage{times}
\usepackage{graphicx}
\usepackage[font={small}]{caption}
\usepackage{subcaption}
\usepackage{amsmath}
\usepackage{amssymb}
\usepackage{tabularx}
\usepackage{placeins}
\usepackage{multirow}
\usepackage{braket}
\usepackage{color}
\usepackage{hyperref}
\usepackage[normalem]{ulem}
\usepackage{breakcites}
\usepackage{adjustbox}
\usepackage{mathtools}
\usepackage{float}
\usepackage{adjustbox}
\usepackage{dcolumn}
\usepackage{bm}
\usepackage{comment}
\usepackage{titlesec}
\usepackage{booktabs}
\usepackage{siunitx}
\usepackage{comment}
\usepackage{url}

\usepackage[dvipsnames]{xcolor}

\titleformat{\subsubsection}{\normalfont\centering}{\thesubsubsection.}{1em}{}
\begin{document}
\title{Accurate atomic correlation and total energies for correlation consistent effective core potentials II: Rb$-$Xe elements} 
\author{Aqsa Shaikh}
\altaffiliation{A. Shaikh and O. Madany contributed equally to this work. \\ Email: ashaikh3@ncsu.edu, osmadany@ncsu.edu} 

\author{Omar Madany}
\altaffiliation{A. Shaikh and O. Madany contributed equally to this work. \\ Email: ashaikh3@ncsu.edu, osmadany@ncsu.edu} 

\author{Benjamin Kincaid}
\author{Lubos Mitas}
\affiliation{Department of Physics, North Carolina State University, Raleigh, North Carolina 27695-8202, USA}
\begin{abstract}{\centering}
We employ correlation-consistent effective core potentials (ccECPs) to perform exact or nearly exact correlation and total energy calculations for the fifth-row elements (Rb-Xe). Total energies are calculated using various correlated methods: configuration interaction (CI), coupled-cluster (CC) up to perturbative quadruple excitations whenever feasible, and stochastic quantum Monte Carlo (QMC) approaches. In order to estimate the energy at the complete basis set (CBS) limit, the basis sets are constructed systematically through aug-cc-p(C)VnZ for each ccECP and further extrapolated to the CBS limit within the corresponding methods. Kinetic energies are evaluated at the FCI/CISD level to provide insights into the electron density and localization of the ccECPs. We also provide data sets for widely used diffusion Monte Carlo (DMC) to quantify fixed-node biases with single-reference trial wavefunctions. These comprehensive benchmarks validate the accuracy of ccECPs within the CC, CI, and QMC methodologies, thus providing accurate and tested valence-only Hamiltonians for many-body electronic structure calculations. 
\end{abstract}
\maketitle
\input{intro.tex}
\input{methods.tex}
\input{results/results.tex}
\input{conclusions.tex}

\section{Associated Content}
\label{sec: suppl_data}
We provide additional useful information in the Supplementary document. Which contains extrapolated energies from $D_{2h}$ point group symmetry for all the \textit{ab initio} methods, QMC results with associated variances and FN-DMC energies at each time step. We also provide various checks for robustness of extrapolation schemes employed in the paper. Along with these additional data points, ccECP parameters and aug-cc-p(C)V$n$Z basis sets for elements Technetium (Tc) and Xenon (Xe) are included in the last section of supplementary data.

\section*{Acknowledgments}
The authors thank Paul R. C. Kent and Jaron T. Krogel for reading the manuscript and for providing helpful suggestions.
They are also grateful to Dr. Abdulgani Annaberdiyev for the fruitful discussions.

This work has been supported by the U.S. Department of Energy, Office of Science, Basic Energy Sciences, Materials Sciences and Engineering Division, as part of the Computational Materials Sciences Program and Center for Predictive Simulation of Functional Materials.

This research used resources of the National Energy Research Scientific Computing Center (NERSC), a U.S. Department of Energy Office of Science User Facility operated under Contract No. DE-AC02-05CH11231.

An award of computer time was provided by the Innovative and Novel Computational Impact on Theory and Experiment (INCITE) program.

This research used resources of the Oak Ridge Leadership Computing Facility, which is a DOE Office of Science User Facility supported under Contract No. DE-AC05-00OR22725.

This paper describes objective technical results and analysis. Any subjective views or opinions that might be expressed in the paper do not necessarily represent the views of the U.S. Department of Energy or the United States Government.

Note:  This manuscript has been authored by UT-Battelle, LLC, under contract DE-AC05-00OR22725 with the US Department of Energy (DOE). The US government retains and the publisher, by accepting the article for publication, acknowledges that the US government retains a nonexclusive, paid-up, irrevocable, worldwide license to publish or reproduce the published form of this manuscript, or allow others to do so, for US government purposes. DOE will provide public access to these results of federally sponsored research in accordance with the DOE Public Access Plan (http://energy.gov/downloads/doe-public-access-plan).

\section*{Conflict of Interest}
The authors have no conflicts to disclose.
\section*{Author Contributions}
\label{sec:auth_contrib}
\textbf{Aqsa Shaikh}: Conceptualization (equal); Data curation (equal); Investigation (equal); Methodology (equal); Visualization (lead); Validation (lead); Writing – original draft (equal); Writing –review \& editing (equal).
\textbf{Omar Madany}: Conceptualization (equal); Data curation (equal); Investigation (equal); Methodology (equal); Visualization (supporting); Validation (lead); Writing – original draft (lead); Writing –review \& editing (equal). 
\textbf{Benjamin Kincaid}: Conceptualization (lead); Data curation (supporting); Investigation (supporting); Methodology (equal); Supervision (supporting); Visualization (supporting); Validation (supporting); Writing – original draft (supporting); Writing –review \& editing (supporting). 
\textbf{Lubos Mitas}: Conceptualization (equal); Investigation (supporting); Methodology (equal); Project administration (lead); Supervision (lead);
Writing – original draft (equal); Writing – review \& editing(equal); Funding acquisition(lead).

\section*{DATA AVAILABILITY}
\label{data}
The data supporting the findings of this study are available in the article and its supplementary material.\\
Additional supporting research data, input and output files for this work, may be accessed through Materials Data Facility.
\section*{REFERENCES}
\bibliographystyle{apsrev4-2}
\bibliography{main.bib}
\end{document}

%% file: intro.tex
\section{Introduction}\label{intro}
The computational cost of accurately solving the electronic structure of multi-electron systems increases with atomic number $(Z)$. This is particularly important  for many-body approaches based on stochastic sampling due to the rapid growth of total energies from tightly bound core electrons.
 The core states nominally dominate the energy scale so that the scaling of a quantum Monte Carlo (QMC) calculation is proportional to $\approx$ $O(Z^6)$ \cite{RMP, QMCscaling2, QMCscaling3}. For heavier atoms, core-valence correlations  can also affect the results obtained in basis set approaches such as Coupled Cluster (CC) and therefore require further elaboration \cite{CCscaling}. 
 More generally, there are additional challenges such as increased complexity of many-body wavefunctions, and impacts from relativity such as spin-orbit in heavier elements that further hamper accurate descriptions. Assuming our interest is in the valence-only properties, it is common to use effective core potentials
(ECPs) (or pseudopotentials) which 
 provide substantial gains not only on the side of reducing $Z\to Z_{\text{eff}}$ due to removed cores but also by smoothing out the states in the core regions.  Hence, ECPs are very valuable in modern electronic structure practice, enabling computationally tractable yet accurate descriptions of valence electronic electronic structure \cite{dolg-rev}.

 Many ECPs were  built on the top of self-consistent methods such as Hartree-Fock (HF), Dirac-Fock and a variety of Density Functional Approximations. 
 These constructions aimed to reproduce one-particle properties such as charge/norm conservation and low-lying atomic spectra \cite{BFD-2007,LANL-1985,CRENBL-1979,SBKJC-1984,SDF-1982,MWB-1993,MDF-2006}. 
However, significant advances in correlated methods such as stochastic configuration interaction approaches, auxiliary field, and real space sampling QMC \cite{advcorr1,advcorr2,QMCPack} created a need for accurate and many-body probed ECPs. For this purpose, we have introduced a new generation of correlation consistent ECPs (ccECPs) by employing many-body constructions and systematic validations across atomic spectra and small molecular systems~\cite{Bennett2017,Bennett2018,Annaberdiyev2018,Wang2019,Wang2022,Kincaid2022,Haihan2024,Omar2025}. 

Since our ccECPs represent a new set of effective valence-only Hamiltonians, it is important to test their properties. In particular, this involves accurate values of total and correlation energies for reference purposes. In addition, accurate values of the kinetic energies also provide electron density estimates, another important characteristic. 
Since ECPs do not conform to the virial theorem, ratios of kinetic to potential (or total) energies vary from atom to atom, depending not only on the number of valence electrons but also on concrete ECP. 
In effect, kinetic energy for every (pseudo)atom is unique  and it provides an average measure of electronic localization inside the core. 
Our prior benchmarks for H$-$Kr elements have provided atomic total and correlation energies for ccECPs with sub-milliHartree accuracy by exploiting widely used quantum chemistry approaches such as Configuration Interaction with single and double excitations (CISD), restricted and unrestricted coupled-cluster singles, doubles, and perturbative triples (R/UCCSD(T)), inclusion of perturbational quadruples CCSDT(Q) as well as full CI (FCI) within feasible computational limits \cite{acc_engI}. Compiling and careful analysis of results from these approaches has enabled us to construct extrapolations to complete basis set (CBS) limits that provide exact or nearly exact (say, within 1\%) estimations of the valence correlation energies.  

Along with quantum chemistry methods we have also provided variational Monte Carlo (VMC) and diffusion Monte Carlo (DMC) energies and variances with single-reference trial wavefunctions. Comparison of DMC energies with CBS results has allowed us to estimate fixed-node (FN) DMC errors \cite{RMP,Reynolds1982,Rasch2015}. Since a significant part of these atomic biases transfers into any multi-atom system, it is particularly important to have clear data in this respect. In addition, trends throughout the periodic table show how errors evolve with the size of the valence spaces, types of occupations, and related considerations.  

This work continues our previous effort and provides a dataset for ccECPs of fifth-row elements (Rb$-$Xe), where relativistic effects and intricate $4d/5p$ electron correlations pose new challenges. For these heavier elements, we employ the same systematic methodology as before. Starting with a state-averaged (SA) Multi Configurational Self Consistent Field (MCSCF) reference state, the orbitals are obtained by optimizing several electronic states of identical spin multiplicity. This averaging enforces invariance under the full rotational symmetry group SO(3), thereby restoring atomic like degeneracies and yielding equivalent orbitals that transform as pure spherical harmonics \cite{state-average}. In the supplementary material, we also provide the same results obtained using the $D_{2h}$ point group, for the irreducible representation corresponding to the lowest energy at the HF level. We note that typically the total energies in both cases remain within systematic uncertainties. This construction is followed by correlation-consistent basis set (aug-cc-p(C)V$n$Z) extrapolations to the complete basis set (CBS) limit to obtain exact (or nearly exact) total energies. More details on the extrapolation methodology are present in section ~\ref{subsec:extrapolation}. We also tested various extrapolation schemes and formulae, results for which are presented in the supplementary material.
In QMC benchmarking, we employ single-determinant trial functions in order to provide values of FN errors relevant for the most commonly used DMC calculations. This data is very useful for future reference and for more detailed analysis, as given below.

 In what follows, our data and analysis focus on the following three aspects: (1) benchmarks of total energies via CCSDT(Q) and FCI within feasible computational limitations extrapolated to the CBS limit;  (2) quantification of fixed-node DMC errors and (3) evaluation of kinetic energy using correlated CI wavefunctions. We performed these analyses for the 5$s$ metals (Rb$-$Sr), 4$d$ transition metals (Y$-$Cd) and 5$p$ main-group elements (In$-$Xe).
~\cite{Bennett2017,Bennett2018,Annaberdiyev2018,Wang2019,Wang2022,Kincaid2022,Haihan2024,Omar2025}, (2) molecular binding curve fidelity ~\cite{Bennett2017,Bennett2018,Annaberdiyev2018,Wang2019,Wang2022,Kincaid2022,Haihan2024,Omar2025}, and (3) fixed-node error in DMC.
Section \ref{sec:methods} details the parameterization of ccECPs, the extrapolation methodology, and the post-HF approaches. The results section \ref{sec:results} is divided into three parts with details of the total energies for Rb--Xe in subsection \ref{subsec:tot_eng_bench}, the analysis of the DMC fixed-node error in subsection \ref{subsec:FN}, and the kinetic energy benchmarks in subsection \ref{subsec:kin_eng_bench}. We conclude our findings in section\ref{conclusion}.

%% file: methods.tex
\section{Methods}
\label{sec:methods}
\subsection{ccECP form}
\label{subsec:ccECP}
The ccECPs used throughout this work are of semi-local form incorporating averaged spin-orbit effects. The averaged relativistic effective potential (AREP) is expressed as $V^{\text{AREP}}_i$:
\begin{equation}\begin{aligned}\label{Varepform}
    V^{\text{AREP}}_i &= V_L(r_i) 
    + \sum_{\ell=0}^{\ell_{\text{max}}=L-1} \left( V_\ell(r_i) - V_L(r_i) \right) \\
    &\quad \times \sum_{m=-\ell}^{\ell} | \ell m \rangle \langle \ell m |,
\end{aligned}\end{equation}
where $r_i$ is the $i^{\text{th}}$ electron's distance from the nucleus, and $V_\ell - V_L$ represents the nonlocal component of $V^{\text{AREP}}_i$. The nonlocal part is defined as:
\begin{equation}\label{eq:AREP_nonlocal}
    V_\ell(r_i) - V_L(r_i) = \sum_{k=1} \beta_{\ell k} r^{n_{\ell k} - 2} e^{-\alpha_{\ell k} r^2}
\end{equation}
with $n_{\ell k}$ as fixed integers (typically $n_{\ell k}=2$ prior to optimization). The local potential $V_L$ takes the form:
\begin{equation}\label{eq:AREP_local}
    V_L (r_i) = - \frac{Z_{\text{eff}}}{r} \left( 1 - e^{\alpha r^2} \right) + \alpha Z_{\text{eff}} r e^{-\beta r^2} + \sum_{i=1}^{2} \gamma_i e^{-\delta_i r^2}    
\end{equation}
ensuring cancellation of the Coulomb singularity as $r \rightarrow 0$. Greek symbols ($\alpha$, $\beta$, $\gamma_i$, $\delta_i$) denote optimized Gaussian parameters, while $Z_{\rm eff} = Z - Z_{\rm core}$ corresponds to the valence electron count. 

In order to estimate the CBS energy for an atomic system, together with ccECPs, we also developed correlation consistent aug-cc-p(C)V$n$Z basis sets, where, $n \in \{D, T, Q, 5\}$) for transition metals and $n \in \{D, T, Q, 5, 6\}$) for main group elements. Following the suit of ccECPs, the construction and optimization of the basis sets is also done within correlated many-body frameworks to ensure correlations are accurately captured at all levels. Further details of ccECPs and their associated correlation-consistent Gaussian basis sets are available in the pseudopotential library and within the associated references\cite{website,Bennett2017,Bennett2018,Annaberdiyev2018,Wang2019,Wang2022,Kincaid2022,Haihan2024,Omar2025}, except for the Tc and Xe AREPs and basis sets, which are provided in the supplementary material.

\subsection{Extrapolation methodology}
\label{subsec:extrapolation}
We calculated the ground-state atomic energies for Rb-Xe elements using ccECPs with correlation-consistent basis sets ((aug)-cc-p(C)V$n$Z, $n \in \{D, T, Q, 5, (6)\}$) across a handful post-HF methods. The HF CBS limit was obtained via the extrapolation scheme\cite{extrapolation}:
\begin{equation}
\label{eq:hfextrap}
    E^{\rm HF}_n = E^{\rm HF}_{\rm CBS} + a \exp\left[ - b n \right]
\end{equation}
while the correlation energy CBS limit was employed:
\begin{equation}
\label{eq:corrextrap}
    E^{\rm corr}_{n} = E^{\rm corr}_{\rm CBS} + \frac{\alpha}{(n+3/8)^{3}} + \frac{\beta}{(n+3/8)^5}
\end{equation}

Here, $a, b, \alpha$ and $\beta$ are fitting parameters, $n$ represents the cardinality of the basis set and $E_{CBS}$ is the final CBS extrapolated energy. In the supplementary material, we evaluated multiple standard extrapolation approaches for both HF and correlation energies\cite{extrapolation}. All HF extrapolation methods tested agreed within our estimated uncertainties. For the correlation energy, our approach (Eq.~\ref{eq:corrextrap}) exhibited general consistency with other techniques; however, the rapid decay of its $(n+3/8)^{-5}$ term systematically underestimates transition metal correlation energies by 2-4 mHa relative to the multi-exponent fit ($E^{\rm corr}_n = E^{\rm corr}_{\text{CBS}} + \sum_i a_i (n+0.5)^{-x_i}$, $x_i \in 3,4$). This conservative approach ensures that our CBS-extrapolated values avoid overestimation biases, providing rigorously bounded and reliable reference data.

It should be noted that in a handful of cases the HF energy shows an increase within sub mili-Hartrees 
while going from lower quality basis set to the higher one, this is a consequence of optimizing the basis set for total energy which includes correlation energy as well. To avoid any discrepancy, we ensured that the extrapolated HF energy is bounded above by the minimum of HF energy obtained for various basis sets. This condition does not affect the uncertainty in the total energy since it is much smaller than the systemic uncertainty in the correlation energy extrapolation.
\subsection{Total energy estimation}
\label{subsec:tot_eng}
\subsubsection{Correlated quantum chemistry methods}
\label{subsubsec:corr}
For all elements and their corresponding ccECPs, we perform the CISD. For open-shell systems, both RCCSD(T) and UCCSD(T) were included. FCI was used for systems with $\leq 6$ valence electrons; otherwise, CCSDT(Q) was applied using \textsc{MOLPRO} and the integrated \textsc{MRCC} package \cite{molpro-2012,molpro-2020,molpro-3,molpro-cc,molpro-seward,molpro-mcscf1,molpro-mcscf2,molpro-fci,MRCC-1,MRCC-2}. 

In systems with two valence electrons (e.g., Sr[\,[Ar]$3d^{10}$]), CISD becomes exact within the given basis set, necessitating only HF and CISD energies. This truncation also applies to systems with $\leq 3$ valence electrons (e.g., In[\,[Kr]$4d^{10}$]), where CCSDT(Q) offers negligible improvement over CCSDT.

For the cases where calculations with $n \in \{5Z~\text{or}~6Z\}$ basis sets were computationally intractable, the missing high-cardinality energies were estimated using a ratio-based approach \cite{acc_engI}. The correlation energy ratio in the largest feasible basis set ($n_{\text{max}}-1$) was calculated as:
\begin{equation}
\label{eq:missing_extrap_ratio}
    \eta_{n_{\text{max}}-1} = \frac{E^{\text{corr,CCSDT(Q)}}_{(n_{\text{max}}-1)}}{E^{\text{corr,UCCSD(T)}}_{(n_{\text{max}}-1)}},
\end{equation}

where $E^{\text{corr}}$ denotes the correlation energy. The missing CCSDT(Q) energy at the basis $n_{\text{max}}$ was then estimated by:
\begin{equation}
\label{eq:missing_extrap_eq}
    E^{\text{corr,CCSDT(Q)}}_{(n_{\text{max}})} = \eta_{n_{\text{max}}-1} \cdot E^{\text{corr,UCCSD(T)}}_{(n_{\text{max}})}.
\end{equation}

Similarly, FCI values that were out of scope for feasibility were estimated from the correlation energy ratio between FCI and CCSDT(Q) results for the largest basis set runs. Although not strictly exact, this procedure proved to be reasonably systematic so as to provide very accurate correlation energy estimates for our purposes, i.e., typically within 1\% of the valence correlation energy. Further validations using computationally feasible cases revealed only marginal systematic errors (see Table S.17 in the supplementary material). This has enabled us to complete the desired estimations for all the considered elements while avoiding prohibitively large calculations with the largest basis sizes in elements with larger valence spaces. 

\subsubsection{Quantum Monte Carlo methods}
\label{subsubsec:dmc}

Fixed-node DMC energies were computed for all atoms. In order to eliminate the time-step bias in the Green's function approximation, total energies were extrapolated to the zero time-step limit ($\tau \to 0$) via linear regression using four discrete steps:
\begin{equation}
    \tau = \{0.02,\ 0.01,\ 0.005,\ 0.0025\}\ \text{Ha}^{-1}.
\end{equation}

The T-moves algorithm\cite{t-moves} ensures the variational evaluation of the nonlocal ccECP AREP component, thus providing a rigorous upper bound to the true energy of the ground state. The trial wavefunction $\Psi_T$ is comprised of a single-reference Slater determinant 
multiplied by explicit Jastrow correlation factors:
\begin{equation}\label{eq:trialwf}
    \Psi_T = \Phi_{\text{SR}} \cdot J_{\text{eI}} \cdot J_{\text{ee}} \cdot J_{\text{eeI}},
\end{equation}
where$J_{\text{eI}}$,$J_{\text{ee}}$, and $J_{\text{eeI}}$ denote one-body electron-ion, two-body electron-electron, and three-body electron-electron-ion terms, respectively.

The initial single Slater determinant was (SD) generated at the HF level using \textsc{Gamess}\cite{GAMESS} or \textsc{PySCF}\cite{PYSCF} at $D_{2h}$ point group with augmented $TZ/QZ$ level basis, as per our previous conclusion that basis set quality has no significant effect on single reference DMC results due to HF nodes being correctly captured at $DZ$ level for atomic systems \cite{acc_engI}. After the construction of slater determinant, the Jastrow factor optimization followed a hierarchical procedure: (1) Optimization of one-body ($J_{\text{eI}}$) and two-body ($J_{\text{ee}}$) terms. (2) Final optimization incorporating three-body ($J_{\text{eeI}}$) correlations. This sequential approach ensures a systematic reduction in the variance of the total energy. All VMC and DMC calculations utilized the \textsc{QMCPACK}\cite{QMCPack} software.
\subsection{Kinetic energy estimation}
\label{subsec:kin_eng}
Accurate kinetic energy evaluation plays an important role for two primary reasons: (1) the virial theorem is inapplicable to ECPs due to modified valence orbital shapes and absent core states; (2) The role of kinetic energy as a metric for electron density spatial localization is particularly relevant in QMC where Jastrow optimizations alter the electron densities. These modifications, most pronounced when seeking the balance of accuracy between  nuclei and in tail regions, arise from variational energy being less sensitive to wavefunction tail behavior. We note that significant deviations from the reference kinetic energies indicate electronic density biases together with suboptimal wavefunction optimizations.

Atomic kinetic energies are given as customary:
\begin{equation}
    E_{\text{kin}} = -\frac{1}{2} \frac{\langle \Psi | \nabla^2 | \Psi \rangle}{\langle \Psi | \Psi \rangle},
    \label{kinetic_expression}
\end{equation}
where $\Psi$ denote the CISD/FCI wavefunctions.  The CBS limit was estimated via a two-point scheme:
\begin{equation}
    E^{\rm kin}_{\rm CBS} = E^{\rm kin}_{\rm n_{\text{max}}} + (E^{\rm kin}_{\rm n_{\text{max}}} - E^{\rm kin}_{\rm n_{\text{max}}-1}),
    \label{kinetic_CBS}
\end{equation}
with extrapolation error:
\begin{equation}
    \sigma = \frac{|E^{\rm kin}_{\rm n_{\text{max}}} - E^{\rm kin}_{\rm n_{\text{max}}-1}|}{2},
\end{equation}
Transition metals used $n_{\text{max}}=5$, while main group elements employed $n_{\text{max}}=6$. 
 Kinetic energy estimates from QMC methods are also presented in the supplementary data (Table S.5) that covers time step extrapolation  obtained from SD-DMC and VMC with and without optimized three body Jastrows factors. 

%% file: results/results.tex
\section{Results}
\label{sec:results}
Our systematic analysis of fifth-row ccECP performance spans three domains: (1) Total energy benchmarks that establish reference values converged with CBS extrapolations (Subsection~\ref{subsec:tot_eng_bench}), (2) quantification of the FN-DMC error across core approximations (Subsection~\ref{subsec:FN}), and (3) kinetic energy evaluations probing pseudopotential-induced density modifications (Subsection~\ref{subsec:kin_eng_bench}).
\subsection{Total energy benchmarks}
\label{subsec:tot_eng_bench}
Accurate atomic total energies were obtained via basis set extrapolation (Eqs.~\eqref{eq:hfextrap}--\eqref{eq:missing_extrap_eq}) and method-specific hierarchies outlined in Subsection~\ref{subsubsec:corr}. State-averaged results are presented below.

The atomic systems employed three different ECP core sizes: 
\begin{itemize}
    \item Rb--Sr used a [Ar]$3d^{10}$ core with aug-cc-pCV$n$Z ($n = 2$--6) basis sets, where the CBS limit was extrapolated using TZ--6Z energies (Table~\ref{tab:tot_Rb-Sr28_ccECPs}).
    \item Y--In also used a [Ar]$3d^{10}$ core with aug-cc-pCV$n$Z ($n = 2$--5) basis sets (Tables~\ref{tab:tot_Y-Tc_ccECPs}--\ref{tab:tot_Ru-In_ccECPs}).
    \item Table \ref{tab:tot_Sr36-Xe_ccECPs} includes:
    \begin{itemize}
        \item Sr with a [Kr] core and aug-cc-pCV$n$Z ($n = 2$--6) basis sets (CBS: TZ--6Z)
        \item In--Xe with a [Kr]$4d^{10}$ core and aug-cc-pV$n$Z ($n = 2$--6) basis sets (CBS: TZ--6Z).
    \end{itemize}
\end{itemize}

The energies computed with the $D_{2h}$ point group symmetry are provided in the supplementary material (Table S.1--S.4). The total energies of the state averaged and $D_{2h}$ calculations agree within $\approx$1 mHa.

\input{results/tot_eng_tables}

\subsection{Fixed-node DMC biases and locality errors}
\label{subsec:FN}
\begin{figure}[htbp!]
\includegraphics[width=\columnwidth]{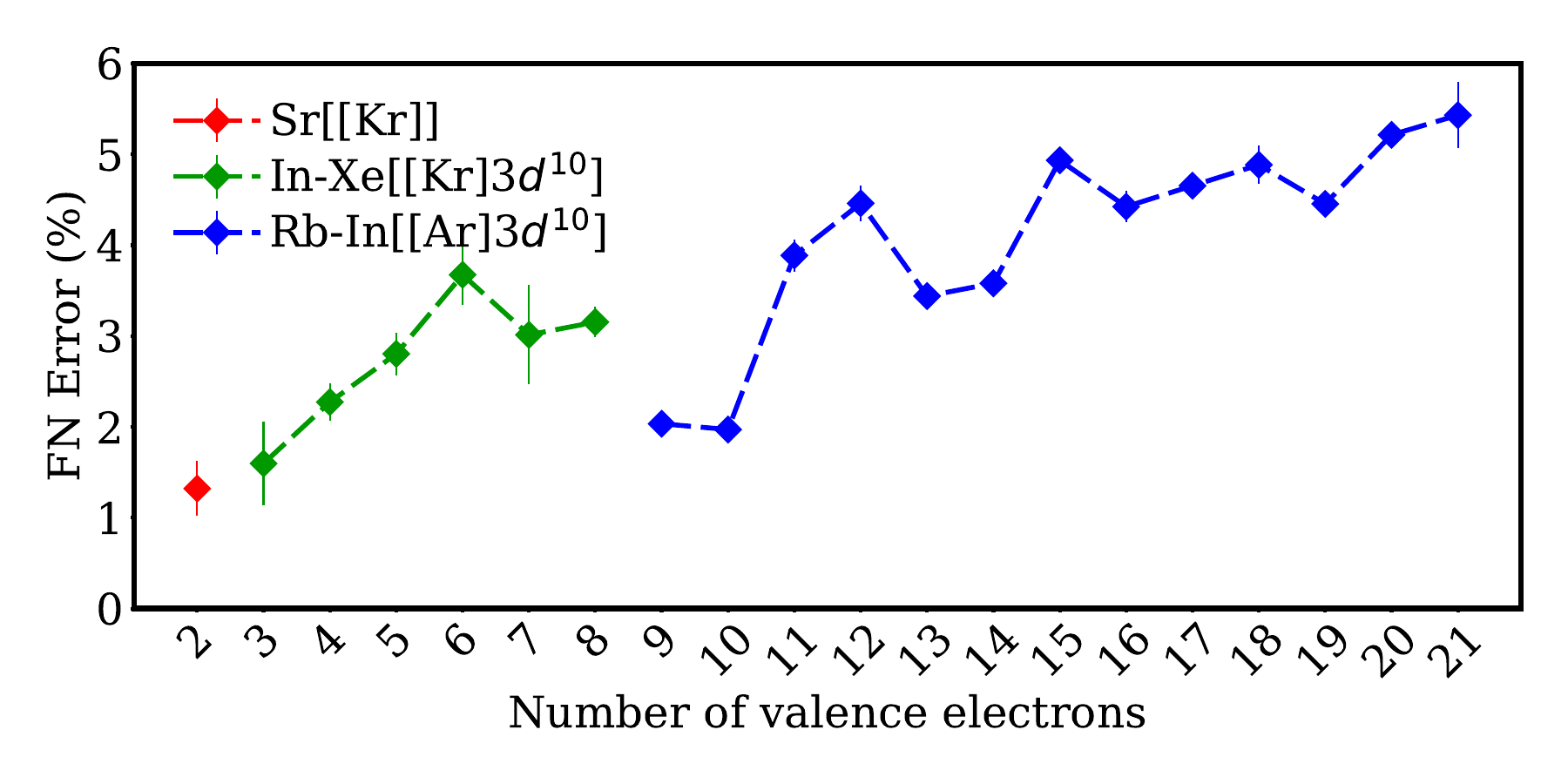}
\caption{Fixed-node DMC errors ($\epsilon$) for ccECPs, as a percentage of the correlation energy using single-reference trial functions: $100\epsilon/|E_{corr}|$. Labels indicate the core sizes employed for each element set in the current study. T-moves were used in all the calculations.}
\label{fig:FN-all}
\centering
\end{figure}

\begin{figure}[htbp!]
\includegraphics[width=\columnwidth]{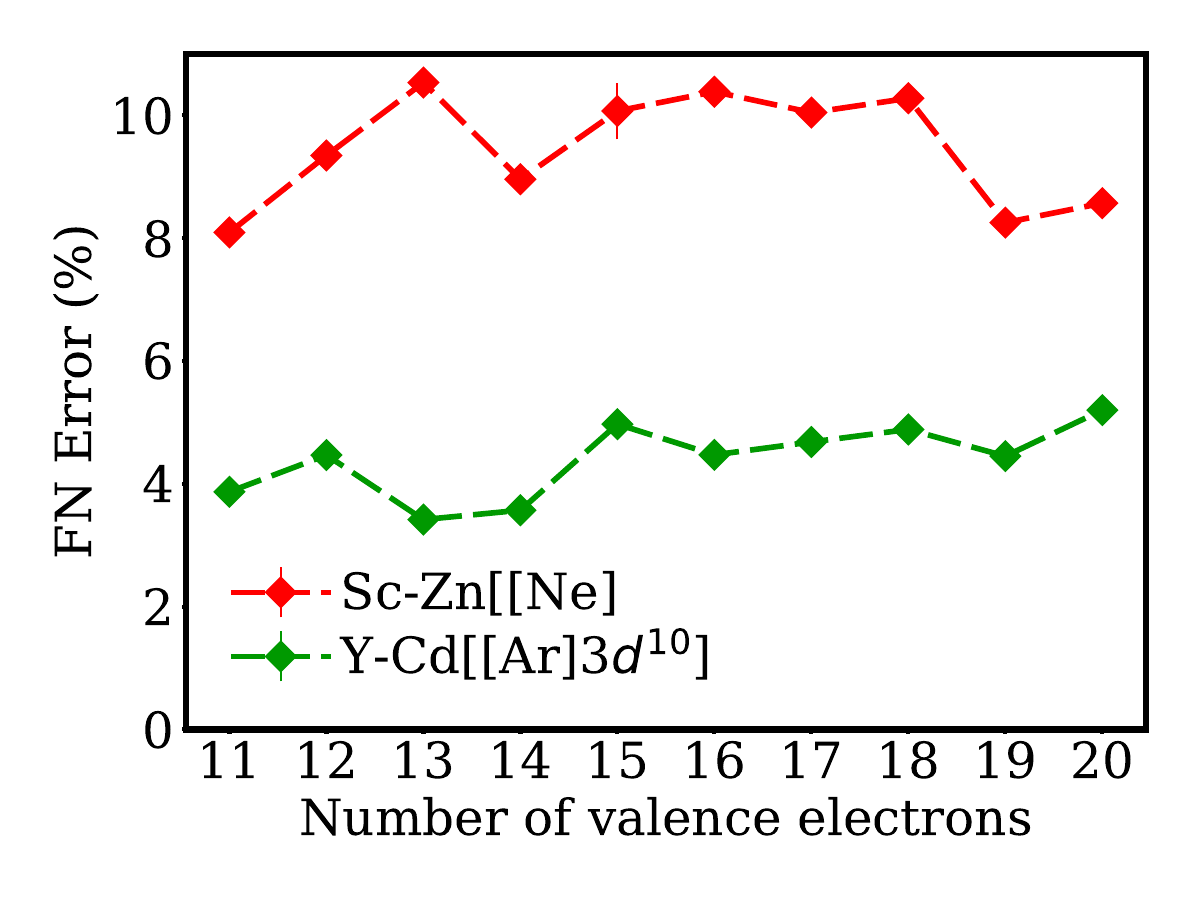}
\caption{Fixed-node DMC errors ($\epsilon$) for ccECPs, as a percentage of the correlation energy using single-reference trial functions: $100\epsilon/|E_{corr}|$. Comparison is shown between $3d$ and $4d$ transition metals. Data for $3d$-TM are from our previous work \cite{acc_engI}}
\label{fig:FN-3d-4d}
\centering
\end{figure}

Systematic DMC/ECP errors arise from FN approximations and nonlocal pseudopotential localization as shown in figure \ref{fig:FN-all} as well as tables\ref{tab:FN-Rb-In28} and \ref{tab:FN-Sr36-Xe}. We point out that the FN biases amplify when the valence angular momentum channel ($p$, $d$ ,...) becomes occupied for the first time (e.g., in $2p$ elements, $3d$ elements, etc). Specifically, 4$d$ transition metals display smaller nodal errors than their 3$d$ counterparts as shown in figure \ref{fig:FN-3d-4d}, in analogy to the 2$p$ vs. 3$p$ pattern in their corresponding rows\cite{acc_engI}. This originates from higher densities and localization of $3d$-orbitals in the core region. In contrast,   4$d$ states are being pushed out by occupied $3d$ levels in the core, leading to smoother and less localized character. That makes the nodal shapes smoother and less curved. Consequently, their accurate description is easier and that leads to a decrease in  the nodal biases.  Figure ~\ref{fig:FN-3d-4d} shows this trend with a clear reduction of the FN error in 4$d$ transition metal
atoms. This is important in general since this part of the atomic FN biases is almost fully transferred to larger systems and mostly cancels out in the energy differences.  

The locality error that has to do with the projection of the nonlocal operator onto the trial function is lumped into the fixed-node bias and is usually difficult to disentangle. As mentioned above,
throughout the FN-DMC calculations we use T-moves to keep the upper bound property of the calculated estimations.
Obviously, for Sr  with [Kr] core the nodeless 5$s$ pseudo-orbitals in $5s^2$ ($^1S$) states eliminate the FN errors since the pseudized state wavefunctions are non-negative everywhere. That provides a glimpse of genuine locality error that is indeed very small, $\approx$ 1-1.6 \%  of the correlation energy. We also note that the
ECP localization errors are absent in purely local pseudopotentials (e.g., pseudo-Hamiltonian formulations \cite{PH-David-1989, PH-Reboredo-2023}

\begin{figure}[htbp!]
\includegraphics[width=\columnwidth]{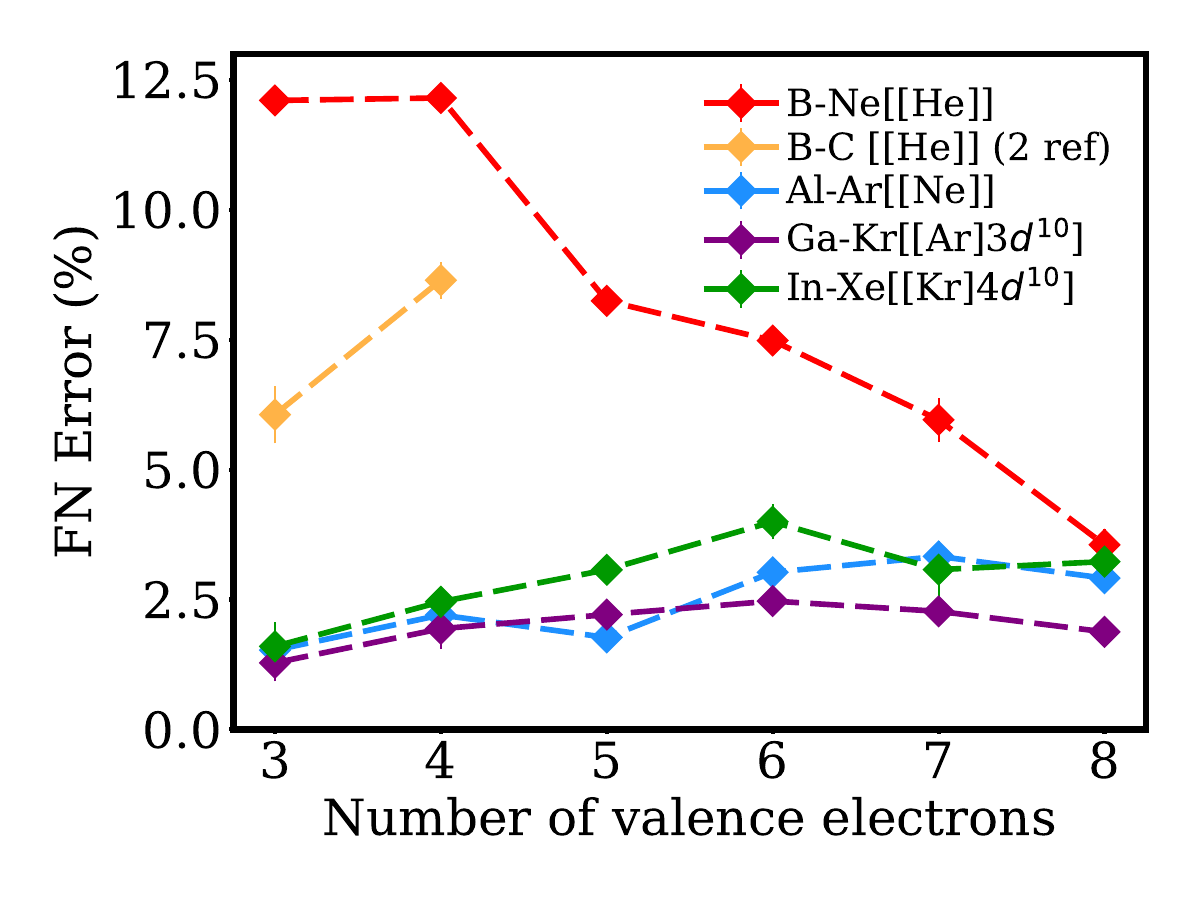}
\caption{Fixed-node DMC errors ($\epsilon$) for ccECPs, as a percentage of the correlation energy using single-reference trial functions: $100\epsilon/|E_{corr}|$. This graph compares FN errors for different rows of p-block main group elements. For B and C atoms we include results with two-reference trial functions that eliminate the pronounced near-degeneracy effects.  Data for $5p$ elements is from current work, the rest is shown for comparison using our previous work \cite{acc_engI}.}
\label{fig:FN-all-p}
\centering
\end{figure}

\begin{figure}[htbp!]
\includegraphics[width=\columnwidth]{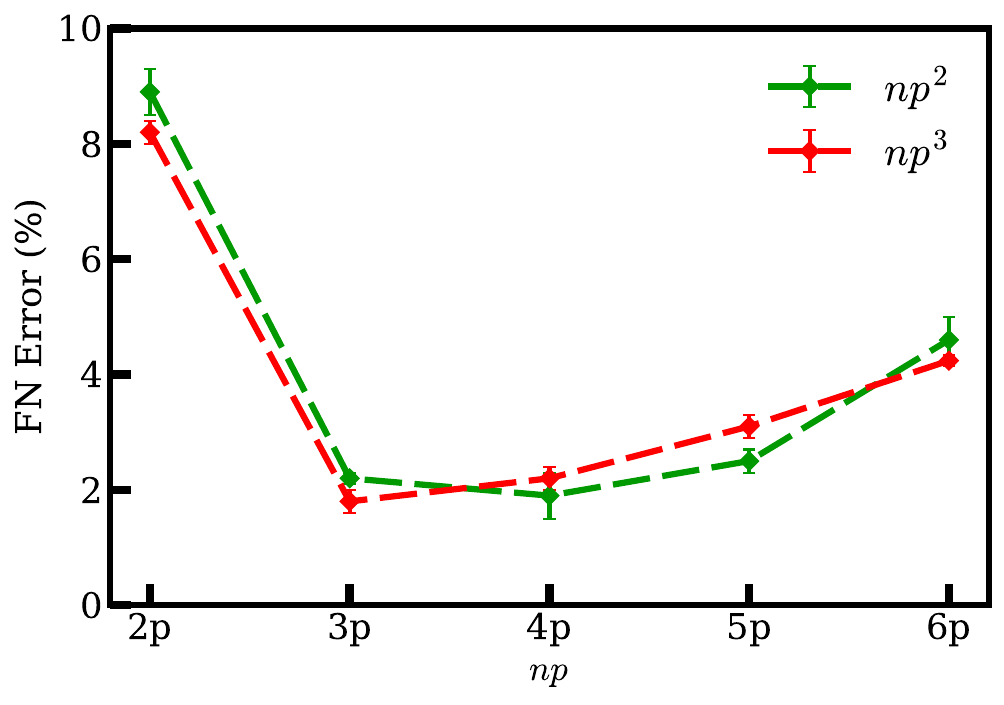}
\caption{Fixed-node DMC errors ($\epsilon$) for ccECPs, as a percentage of the correlation energy using single-reference trial functions: $100\epsilon/|E_{corr}|$. 
All the elements belong to isovalent  $np^2$ and $np^3$ configurations.}
\label{fig: np2 np3}
\centering
\end{figure}

A clear reduction of the FN error occurs for niobium (Nb) (5$s$4$d^4$) when compared to zirconium (Zr) (5$s^2$4$d^2$), mirroring a corresponding effect observed for 3$d$ chromium (Cr) vs vanadium (V) atoms. The FN error decrease of $\sim$1.1\% from Zr to Nb reflects the high-spin electron configuration of Nb in the $^6D$ ground state, which is shown in figure \ref{fig:FN-3d-4d}. High-spin symmetry suppresses configurational mixing via maximized $d$-shell spin alignment, thus minimizing unlike-spin pair correlations that dominate the dynamical correlation effects. Consequently, single-reference HF wavefunctions provide a higher nodal accuracy in Nb than in Zr. These results extend our 3$d$-element findings\cite{acc_engI}, confirming that the half-filled $d$-shell systems universally mitigate FN errors through symmetry-enforced  restrictions on the mixing of excited configurations.
Moving on to $p$-block elements, figures ~\ref{fig:FN-all-p} and ~\ref{fig: np2 np3} compare the trend of FN-biases changing over rows and columns. The completely filled shells of $np^6$ elements have comparably lower errors across all rows despite $2s$ and $2p$ configurations mixing, which reduces significantly with increasing p-occupancy. Figure \ref{fig: np2 np3} compares FN errors across isovalent $np^2$ to $np^3$ elements, 
where, a close comparison between the isovalent $np^2$ and $np^3$ elements shows that the FN errors are greatly reduced as one moves from $2p$ to $3p$, due to the presence of a $p$ shell inside the core. This is despite the fact that for $2p$ the number of electrons in this channel is below the half-occupancy, and we use a two-reference trial wavefunction in order to eliminate well-known (and special) strong near degeneracy effect \cite{Rasch2012}. Still, the FN errors are 3 to 4 times larger than in the other isovalent elements. We also note a small but steady increase in the FN error with the size of the core, which is expected due to the counter-intuitive increase in the wavefunction shape complexity and mixing of the locality error inside the core from strongly repulsive ccECP potentials. 
There are only marginal changes when moving down the periodic table for the main group elements, as observed in Fig. \ref{fig: np2 np3}.
\input{results/fn_bias_tables}   
%


\subsection{Kinetic energies}
\label{subsec:kin_eng_bench}
\begin{figure}[htbp!]
\includegraphics[width=\columnwidth]{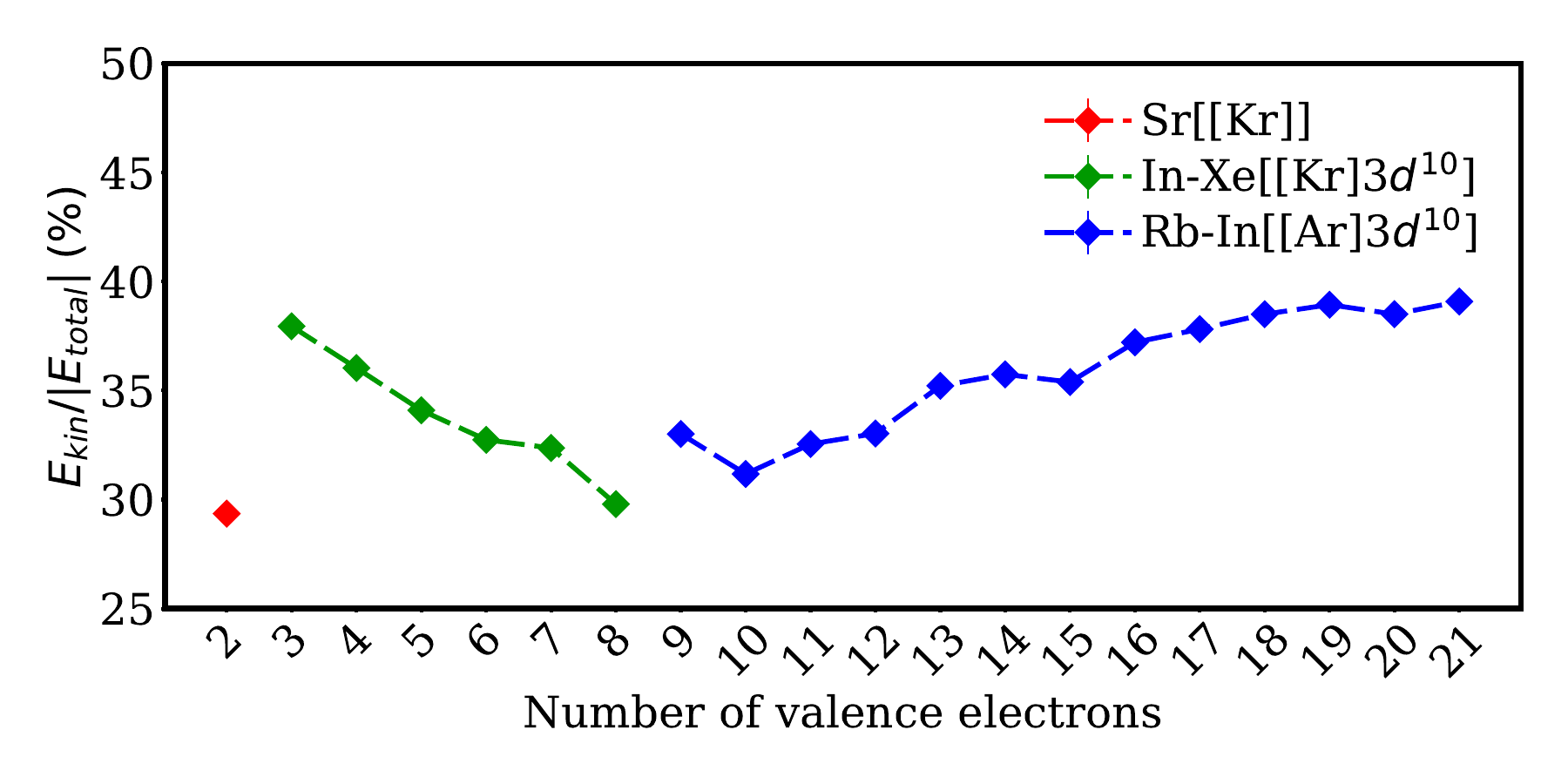}
\caption{Estimated kinetic energy of ccECPs as a percentage of the total energy, $100E_{kin}/|E_{total}|$.
}
\label{fig:Kin-all}
\centering
\end{figure}

\begin{figure}[htbp!]
\includegraphics[width=\columnwidth]{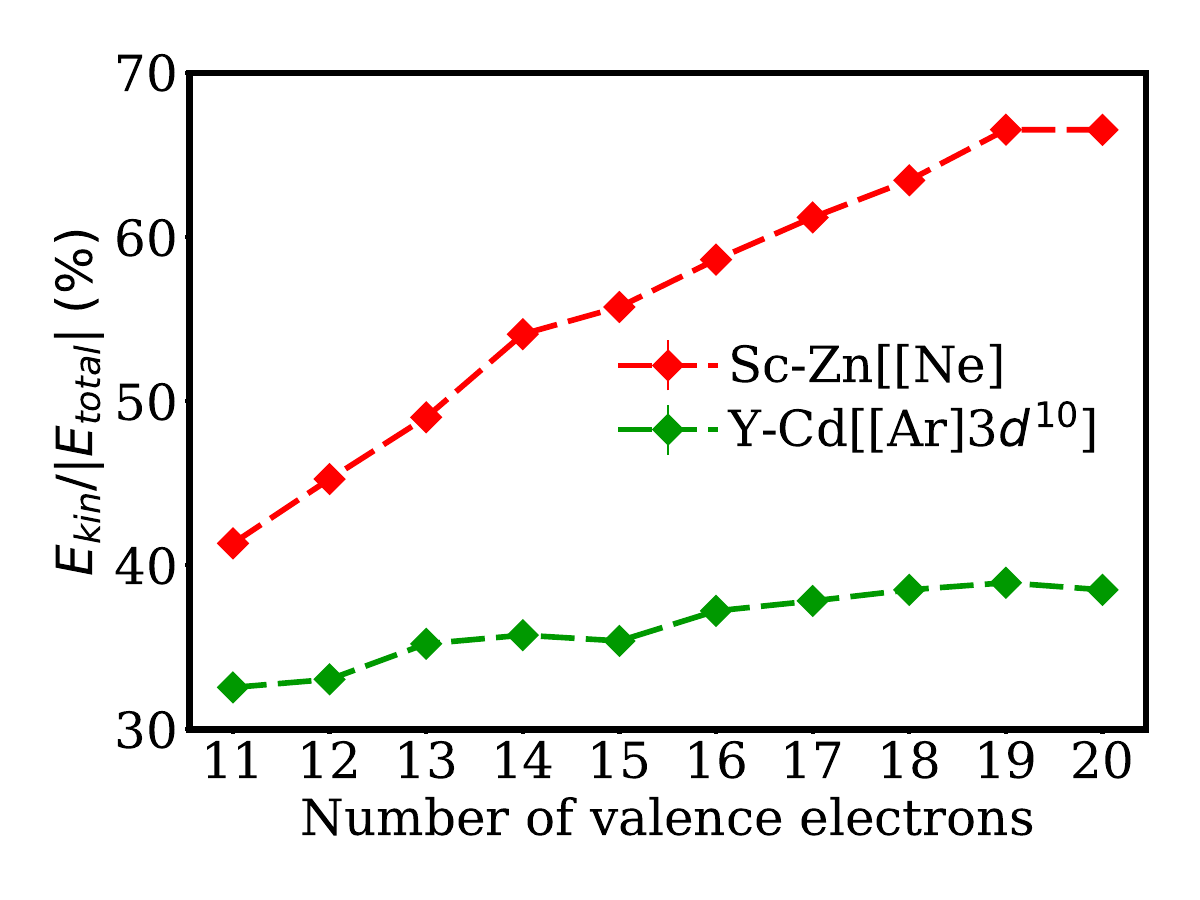}
\caption{Estimated kinetic energy of ccECPs as a percentage of the total energy, $100E_{kin}/|E_{total}|$. Comparing the trends in $3d$ vs $4d$ transition metals.
}
\label{fig:KE-3d-4d}
\centering
\end{figure}

\begin{figure}[htbp!]
\includegraphics[width=\columnwidth]{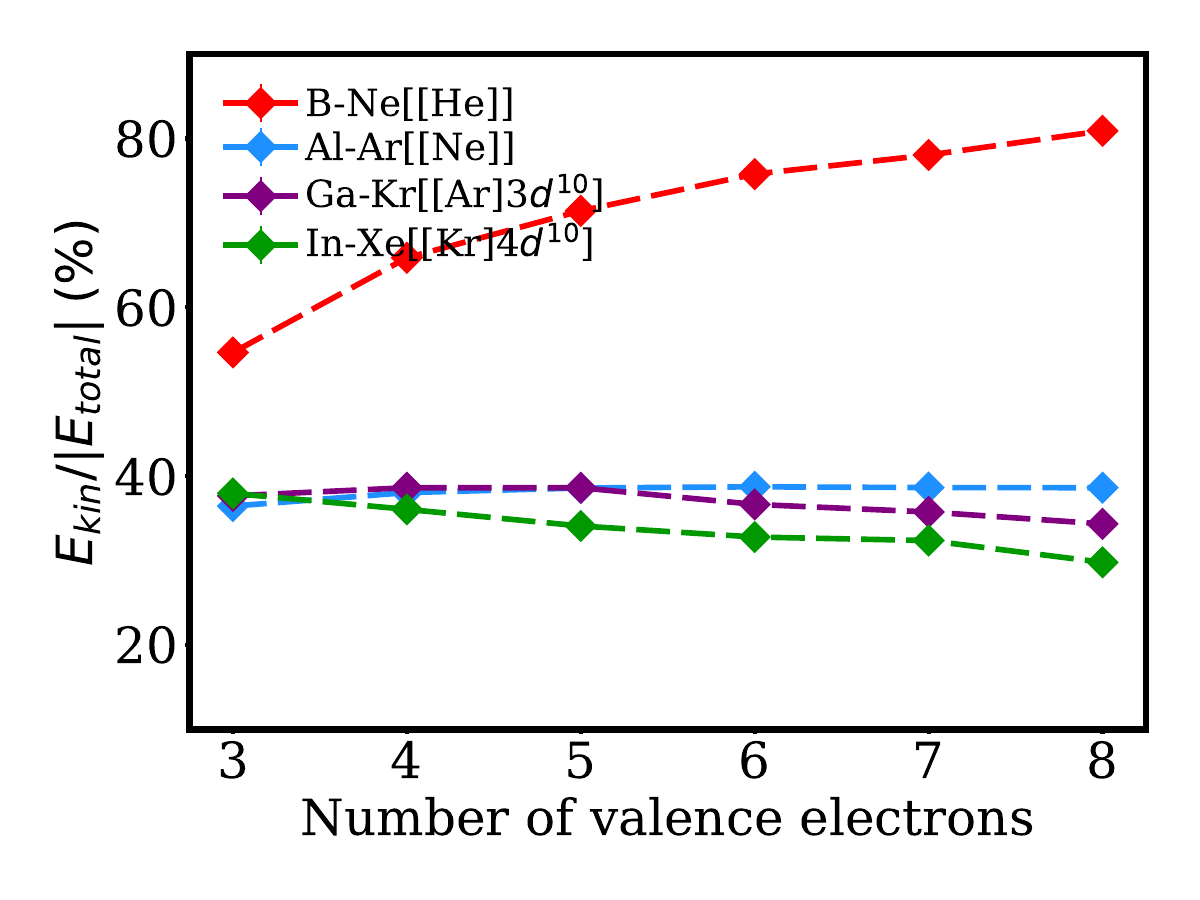}
\caption{Estimated kinetic energy of ccECPs as a percentage of the total energy, $100E_{kin}/|E_{total}|$ for all the $p$-block elements studied so far. 
}
\label{fig:KE-all-p}
\centering
\end{figure}

Kinetic energies for fifth-row ccECP systems (Rb--Xe) follow trends analogous to lighter elements, modulated by relativistic effects and 5$p$/4$d$ orbital hybridization. Tabulated values for Rb--In ([Kr]4$d^{10}$ core), In--Xe ([Kr]4$d^{10}$ core), and Sr ([Kr] core) appear in Tables~\ref{tab:kin_Rb-Mo_ccECPs}--\ref{tab:kin_Sr36-Xe_ccECPs}. Figure~\ref{fig:Kin-all} plots the kinetic-to-total energy ratio versus valence electron count which are also tabulated in \ref{tab:KE-Rb-In28} and \ref{tab:KE-Sr-Xe}. Due the need to use a less accurate approach for larger valence spaces, namely CISD, our kinetic energies for atoms with a larger number of valence electrons are probably underestimated mildly. This is indicated also by comparison with FN-DMC kinetic energies, that are in most cases higher by $\approx$ 2 to 5 \%, please see the supplementary material. Nevertheless, we plot the CBS extrapolated CISD (or FCI whenever feasible) values because FN-DMC provides only mixed-estimators that might carry their own bias \cite{RMP}. We provide a comparative ratio of kinetic energies to total energies for $3d$ vs $4d$ and among main group elements in figures \ref{fig:KE-3d-4d} and \ref{fig:KE-all-p}. 
\input{results/kinetic_tables.tex}

%% file: results/tot_eng_tables.tex
\begin{table*}[htbp!]
\setlength{\tabcolsep}{4pt} 
\centering
\small
\caption{
Total and atomic correlation energies (in Ha) for 5$s$ main group elements (Rb--Sr) calculated using ccECPs with a [[Ar]3$d^{10}]$ core and aug-cc-pCVnZ basis sets. CBS extrapolation were preformed using TZ--6Z levels. Estimated values (marked with *) were obtained from adjacent data points using Eqs.~\ref{eq:missing_extrap_ratio}-\ref{eq:missing_extrap_eq}.
}

\label{tab:tot_Rb-Sr28_ccECPs}
\begin{tabular}{|l|l|ccccc|r|}
\hline\hline
Atom & Method &          $E_{\text{DZ}}$ &          $E_{\text{TZ}}$ &          $E_{\text{QZ}}$ &          $E_{\text{5Z}}$ &          $E_{\text{6Z}}$ &             $E_{\rm CBS}$ \\
\hline
\multirow{4}{*}{\bf{Rb}}
&CISD     &  -0.15794818 &  -0.21383522 &  -0.23403682 &  -0.23992518 &  -0.24234910 &    -0.24506(15) \\
&RCCSD(T) &  -0.16655010 &  -0.23021761 &  -0.25263103 &  -0.25918873 &  -0.26185271 &    -0.26487(12) \\
&UCCSD(T) &  -0.16656447 &  -0.23023455 &  -0.25264741 &  -0.25920493 &  -0.26186888 &    -0.26489(12) \\
&CCSDT(Q) &  -0.16701586 &  -0.23075792 &  -0.25318910 &  -0.25973640 &  -0.26240581(*) &    -0.26542(14) \\
&ROHF      & -23.83664518 & -23.83664981 & -23.83665018 & -23.83665162 & -23.83665672 &  -23.836659(10) \\
\cline{8-8}
&Total    &              &              &              &              &              &   -24.10207(14) \\
\hline
\multirow{4}{*}{\bf{Sr}}
& CISD & -0.22621170 & -0.26804800 & -0.27324308 & -0.27604232 & -0.27733959 & -0.27975(20) \\
& UCCSD(T) & -0.24978960 & -0.29844309 & -0.30420550 & -0.30727702 & -0.30875546 & -0.31142(16) \\
& CCSDT(Q) & -0.25028737 & -0.29901444 & -0.30477778 & -0.30785508(*) & -0.30933630(*) & -0.31201(16) \\
& RHF & -30.21685579 & -30.21685579 & -30.21685579 & -30.21685579 & -30.21685580 & -30.21685580(10) \\
\cline{8-8}
& Total &  &  &  &  &  & -30.52887(16) \\
\hline\hline
\end{tabular}
\end{table*}
\begin{table*}[htbp!]
\setlength{\tabcolsep}{3.5pt} 
\centering
\small
\caption{Total and atomic correlation energies (in Ha) for 4$d$ transition metals (Y--Tc). CBS extrapolation were preformed using DZ--5Z levels. All other methodological details follow Table~\ref{tab:tot_Rb-Sr28_ccECPs}.
}

\label{tab:tot_Y-Tc_ccECPs}
\begin{tabular}{|c|l|cccc|l|}
\hline\hline
Atom & Method &          $E_{\text{DZ}}$ &          $E_{\text{TZ}}$ &          $E_{\text{QZ}}$ &          $E_{\text{5Z}}$ &             $E_{\rm CBS}$ \\
\hline
\multirow{4}{*}{\bf{Y}}
& CISD & -0.26034436 & -0.31017370 & -0.32293639 & -0.32834408 & -0.33375(54) \\
& RCCSD(T) & -0.29348152 & -0.35104961 & -0.36571763 & -0.37184548 & -0.37801(56) \\
& UCCSD(T) & -0.29367684 & -0.35129261 & -0.36597183 & -0.37210210 & -0.37827(56) \\
& CCSDT(Q) & -0.29527850 & -0.35293798 & -0.36761780 & -0.37377564(*) & -0.37994(59) \\
& ROHF & -37.81414061 & -37.81414340 & -37.81414462 & -37.81414511 & -37.814145501(37) \\
\cline{7-7}
& Total &  &  &  &  & -38.19409(59) \\
\hline
\multirow{4}{*}{\bf{Zr}}
& CISD & -0.29214389 & -0.35012600 & -0.36545536 & -0.37175979 & -0.37838(43) \\
& RCCSD(T) & -0.32983600 & -0.39713342 & -0.41470741 & -0.42186883 & -0.42937(47) \\
& UCCSD(T) & -0.33005798 & -0.39742140 & -0.41500700 & -0.42217417 & -0.42968(47) \\
& CCSDT(Q) & -0.33215751 & -0.39945302 & -0.41699882 & -0.42420039(*) & -0.43169(52) \\
& ROHF & -46.34205717 & -46.34206356 & -46.34206544 & -46.34205874 & -46.3420654(63) \\
\cline{7-7}
& Total &  &  &  &  & -46.77375(52) \\
\hline
\multirow{4}{*}{\bf{Nb}}
& CISD & -0.31628033 & -0.38217496 & -0.40215012 & -0.41045121 & -0.41999(27) \\
& RCCSD(T) & -0.34693734 & -0.42253312 & -0.44517362 & -0.45450011 & -0.46521(27) \\
& UCCSD(T) & -0.34715435 & -0.42283870 & -0.44550035 & -0.45483415 & -0.46555(27) \\
& CCSDT(Q) & -0.34795161 & -0.42356013 & -0.44620592 & -0.45555450(*) & -0.46627(29) \\
& ROHF & -56.32326891 & -56.32334234 & -56.32334723 & -56.32334804 & -56.32334841(64) \\
\cline{7-7}
& Total &  &  &  &  & -56.78962(29) \\
\hline
\multirow{4}{*}{\bf{Mo}}
& CISD & -0.34519993 & -0.42035342 & -0.44429727 & -0.45428324 & -0.46608(21) \\
& RCCSD(T) & -0.37868757 & -0.46503934 & -0.49217138 & -0.50342806 & -0.51667(23) \\
& UCCSD(T) & -0.37891628 & -0.46536279 & -0.49252229 & -0.50379014 & -0.51705(23) \\
& CCSDT(Q) & -0.37951535 & -0.46579524 & -0.49294353 & -0.50422102(*) & -0.51749(24) \\
& ROHF & -67.46809562 & -67.46809589 & -67.46809702 & -67.46809749 & -67.4680975(13) \\
\cline{7-7}
& Total &  &  &  &  & -67.98559(24) \\
\hline
\multirow{4}{*}{\bf{Tc}}
& CISD     &  -0.37690775 &  -0.45702407 &  -0.48556478 &  -0.49716136 &      -0.51203(35) \\
& RCCSD(T) &  -0.42375780 &  -0.51721021 &  -0.54982265 &  -0.56312033 &      -0.57995(29) \\
& UCCSD(T) &  -0.42423139 &  -0.51777330 &  -0.55041174 &  -0.56371904 &      -0.58056(29) \\
& CCSDT(Q) &  -0.42522935 &  -0.51842869 &  -0.55100052 &  -0.56432205(*) &      -0.58115(26) \\
& ROHF      & -79.73546552 & -79.73546606 & -79.73546671 & -79.73546675 &  -79.73546685(21) \\
\cline{7-7}
& Total    &              &              &              &              &     -80.31662(26) \\
\hline
\hline\hline
\end{tabular}
\end{table*}
\begin{table*}[htbp!]
\setlength{\tabcolsep}{4pt} 
\centering
\small
\caption{Total and atomic correlation energies (in Ha) for 4$d$ transition metals (Ru--Cd) and 5$p$ main group element (In) for ccECPs with [[Ar]3$d^{10}]$ core. CBS extrapolation were preformed using DZ--5Z levels. All other methodological details follow Table~\ref{tab:tot_Rb-Sr28_ccECPs}.
}
\label{tab:tot_Ru-In_ccECPs}
\begin{tabular}{|l|l|cccc|r|}
\hline\hline
Atom & Method &          $E_{\text{DZ}}$ &          $E_{\text{TZ}}$ &          $E_{\text{QZ}}$ &          $E_{\text{5Z}}$ &             $E_{\rm CBS}$ \\
\hline
\multirow{4}{*}{\bf{Ru}}
& CISD & -0.42549342 & -0.52532506 & -0.56199805 & -0.57674337 & -0.59611(69) \\
& RCCSD(T) & -0.47291288 & -0.58904695 & -0.63122040 & -0.64795226 & -0.67005(92) \\
& UCCSD(T) & -0.47315927 & -0.58940072 & -0.63161862 & -0.64836235 & -0.67048(92) \\
& CCSDT(Q) & -0.47374461 & -0.58972818 & -0.63190616 & -0.64865751 & -0.67078(91) \\
& ROHF & -93.82235798 & -93.82242458 & -93.82243348 & -93.82243911(*) & -93.8224391(28) \\
\cline{7-7}
& Total &  &  &  &  & -94.49322(91) \\
\hline
\multirow{4}{*}{\bf{Rh}}
& CISD     &   -0.4591629 &   -0.5778568 &   -0.6195349 &   -0.6382216 &      -0.66016(93) \\
& RCCSD(T) &   -0.5117063 &   -0.6499699 &   -0.6976508 &   -0.7186443 &      -0.74344(85) \\
& UCCSD(T) &   -0.5119000 &   -0.6502531 &   -0.6979661 &   -0.7189684 &      -0.74378(85) \\
& CCSDT(Q) &   -0.5121257 &   -0.6501825 &   -0.6978553 &   -0.7188543(*) &      -0.74366(85) \\
& ROHF      & -109.3602381 & -109.3607294 & -109.3607623 & -109.3607681 &  -109.3607694(34) \\
\cline{7-7}
& Total    &              &              &              &              &    -110.10443(85) \\
\hline
\multirow{4}{*}{\bf{Pd}}
& CISD & -0.53501025 & -0.67375744 & -0.72446268 & -0.74755853 & -0.7749(12) \\
& RCCSD(T) & -0.60240161 & -0.76670214 & -0.82637181 & -0.85373296 & -0.8858(16) \\
& UCCSD(T) & -0.60240151 & -0.76670207 & -0.82637141 & -0.85373258 & -0.8858(16) \\
& CCSDT(Q) & -0.60162374 & -0.76543094 & -0.82499074 & -0.85230620(*) & -0.8843(16) \\
& RHF & -126.49839999 & -126.49840031 & -126.49840415 & -126.49841936 & -126.498425(21) \\
\cline{7-7}
& Total &  &  &  &  & -127.3828(16) \\
\hline
\multirow{4}{*}{\bf{Ag}}
&CISD     &   -0.5348991 &   -0.6836735 &   -0.7386458 &   -0.7632059 &     -0.79285(87) \\
&RCCSD(T) &   -0.5965612 &   -0.7704389 &   -0.8335984 &   -0.8613548 &     -0.89504(74) \\
&UCCSD(T) &   -0.5966663 &   -0.7705594 &   -0.8337203 &   -0.8614755 &     -0.89516(74) \\
&CCSDT(Q) &   -0.5961459 &   -0.7694036 &   -0.8325165 &   -0.8602316(*) &     -0.89393(71) \\
&ROHF      & -146.0529193 & -146.0529206 & -146.0529220 & -146.0529252 &  -146.052926(11) \\
\cline{7-7}
&Total    &              &              &              &              &   -146.94686(71) \\
\hline

\multirow{4}{*}{\bf{Cd}}
&CISD     &   -0.5635019 &   -0.7132274 &   -0.7753451 &   -0.8022759 &       -0.83730(35) \\
&RCCSD(T) &   -0.6329035 &   -0.8078357 &   -0.8794236 &   -0.9101991 &       -0.95028(52) \\
&UCCSD(T) &   -0.6329036 &   -0.8078356 &   -0.8794235 &   -0.9101989 &       -0.95028(52) \\
&CCSDT(Q) &   -0.6325473 &   -0.8067345 &   -0.8782802 &   -0.9090156(*) &       -0.94913(56) \\
&RHF      & -166.6824187 & -166.6824189 & -166.6824190 & -166.6824195 &  -166.68241957(55) \\
\cline{7-7}
&Total    &              &              &              &              &     -167.63155(56) \\
\hline
\multirow{7}{*}{\bf{In}}
&CISD     &   -0.5835218 &   -0.7528545 &   -0.8206863 &   -0.8529978 &       -0.8913(21) \\
&RCCSD(T) &   -0.6508148 &   -0.8498080 &   -0.9263763 &   -0.9640944 &       -1.0069(37) \\
&UCCSD(T) &   -0.6508903 &   -0.8499250 &   -0.9265094 &   -0.9642307 &       -1.0071(37) \\
&CCSDT(Q) &   -0.6507287 &   -0.8489279 &   -0.9254225(*) &   -0.9630995(*) &       -1.0059(36) \\
&ROHF      & -189.2302376 & -189.2302379 & -189.2302381 & -189.2302415 &  -189.2302421(55) \\
\cline{7-7}
&Total    &              &              &              &              &     -190.2362(36) \\
\hline
\hline\hline
\end{tabular}
\end{table*}
\begin{table*}[htbp!]
\setlength{\tabcolsep}{4pt} 
\centering
\small
\caption{Total and atomic correlation energies (in Ha) for 5$s$ (Sr with [Kr] core) and 5$p$ (In--Xe with [[Kr]4$d^{10}]$ core) main group elements calculated using ccECPs and aug-cc-pVnZ basis sets. CBS extrapolation employed TZ--6Z series. All other methodological details follow Table~\ref{tab:tot_Rb-Sr28_ccECPs}.
}
\label{tab:tot_Sr36-Xe_ccECPs}
\begin{tabular}{|l|l|ccccc|r|}
\hline\hline
Atom & Method &          $E_{\text{DZ}}$ &          $E_{\text{TZ}}$ &          $E_{\text{QZ}}$ &          $E_{\text{5Z}}$ &          $E_{\text{6Z}}$ &             $E_{\rm CBS}$ \\
\hline
\multirow{4}{*}{\bf{Sr}}
& FCI & -0.02508516 & -0.02627378 & -0.02639811 & -0.02645987 & -0.02646605 & -0.026503(25) \\
& RHF & -0.56807473 & -0.56807505 & -0.56807518 & -0.56807524 & -0.56807547 & -0.56807553(62) \\
\cline{8-8}
& Total &  &  &  &  &  & -0.594578(25) \\
\hline
\multirow{4}{*}{\bf{In}}
& CISD     & -0.03917673 & -0.04284554 & -0.04636770 & -0.04679469 & -0.04696061 &    -0.04667(19) \\
& RCCSD(T) & -0.03975159 & -0.04373738 & -0.04743861 & -0.04790242 & -0.04808205 &    -0.04780(20) \\
& UCCSD(T) & -0.03975892 & -0.04378399 & -0.04748847 & -0.04795199 & -0.04813127 &    -0.04784(20) \\
& FCI      & -0.03995559 & -0.04401421 & -0.04771119 & -0.04815602 & -0.04832630 &    -0.04802(20) \\
& ROHF      & -1.84984364 & -1.84984506 & -1.84984427 & -1.84984511 & -1.84984557 &  -1.8498463(19) \\
\cline{8-8}
& Total    &             &             &             &             &             &    -1.89786(20) \\
\hline
\multirow{4}{*}{\bf{Sn}}
& CISD & -0.05356572 & -0.06316306 & -0.06536129 & -0.06610170 & -0.06637733 & -0.066780(42) \\
& RCCSD(T) & -0.05482554 & -0.06564567 & -0.06813017 & -0.06895946 & -0.06926544 & -0.069710(47) \\
& UCCSD(T) & -0.05484750 & -0.06582742 & -0.06831146 & -0.06913832 & -0.06944298 & -0.069884(47) \\
& CCSDT(Q) & -0.05516899 & -0.06638863 & -0.06880875 & -0.06959289 & -0.06987884 & -0.070280(41) \\
& FCI & -0.05518023 & -0.06639984 & -0.06882049 & -0.06960378 & -0.06988936 & -0.070289(41) \\
& ROHF & -3.27365566 & -3.27365363 & -3.27365400 & -3.27365515 & -3.27365664 & -3.2736576(42) \\
\cline{8-8}
& Total &  &  &  &  &  & -3.343938(42) \\
\hline
\multirow{4}{*}{\bf{Sb}}
& CISD     & -0.06340408 & -0.07967920 & -0.08367584 & -0.08481536 & -0.08542305 &    -0.08602(17) \\
& RCCSD(T) & -0.06503015 & -0.08376763 & -0.08841298 & -0.08971542 & -0.09040177 &    -0.09106(20) \\
& UCCSD(T) & -0.06505433 & -0.08410034 & -0.08874038 & -0.09003956 & -0.09072309 &    -0.09137(19) \\
& CCSDT(Q) & -0.06535485 & -0.08482621 & -0.08942101 & -0.09064912 & -0.09129596 &    -0.09186(20) \\
& FCI      & -0.06538006 & -0.08484306 & -0.08944158 & -0.09066997(*) & -0.09131696(*) &    -0.09188(20) \\
& ROHF      & -5.29879795 & -5.29879609 & -5.29879839 & -5.29879797 & -5.29879968 &  -5.2987998(32) \\
\cline{8-8}
& Total    &             &             &             &             &             &    -5.39066(20) \\
\hline
\multirow{4}{*}{\bf{Te}}
& CISD & -0.07540664 & -0.10905611 & -0.11954256 & -0.12226654 & -0.12360173 & -0.12473(38) \\
& RCCSD(T) & -0.07845361 & -0.11597826 & -0.12777627 & -0.13080209 & -0.13226989 & -0.13354(43) \\
& UCCSD(T) & -0.07850123 & -0.11615925 & -0.12801269 & -0.13104050 & -0.13250762 & -0.13370(43) \\
& CCSDT(Q) & -0.07883137 & -0.11679043 & -0.12868967 & -0.13165066 & -0.13307015 & -0.13416(43) \\
& FCI & -0.07884635 & -0.11680415 & -0.12872340(*) & -0.13176802(*) & -0.13324328(*) & -0.13444(43) \\
& ROHF & -8.00739489 & -8.00739400 & -8.00739445 & -8.00739435 & -8.00739501 & -8.00739504(80) \\
\cline{8-8}
& Total &  &  &  &  &  & -8.14155(43) \\
\hline
\multirow{7}{*}{\bf{I}}
& CISD     &  -0.09101153 &  -0.13560238 &  -0.15925115 &  -0.16408053 &  -0.16617905 &      -0.16682(88) \\
& RCCSD(T) &  -0.09523034 &  -0.14556594 &  -0.17239770 &  -0.17783838 &  -0.18017589 &      -0.18084(99) \\
& UCCSD(T) &  -0.09526973 &  -0.14564696 &  -0.17255644 &  -0.17800297 &  -0.18034034 &      -0.18099(99) \\
& CCSDT(Q) &  -0.09562969 &  -0.14628617 &  -0.17337140 &  -0.17876119 &  -0.18103844 &      -0.18156(99) \\
& FCI      &  -0.09564087 &  -0.14630125 &  -0.17338927(*) &  -0.17877962(*) &  -0.18105710(*) &      -0.18157(99) \\
& ROHF      & -11.21512315 & -11.21512201 & -11.21512334 & -11.21512285 & -11.21512267 &  -11.21512343(73) \\
\cline{8-8}
& Total    &              &              &              &              &              &     -11.39668(99) \\

\hline
\multirow{4}{*}{\bf{Xe}}
& CISD     &  -0.11331133 &  -0.17058828 &  -0.19342784 &  -0.20058095 &  -0.20334515 &      -0.20693(16) \\
& RCCSD(T) &  -0.11909792 &  -0.18467207 &  -0.21060710 &  -0.21870292 &  -0.22180642 &      -0.22582(19) \\
& UCCSD(T) &  -0.11909789 &  -0.18467208 &  -0.21060711 &  -0.21870293 &  -0.22180641 &      -0.22582(19) \\
& CCSDT(Q) &  -0.11950338 &  -0.18529614 &  -0.21132957 &  -0.21940690 &  -0.22245015 &      -0.22639(23) \\
& FCI      &  -0.11951268 &  -0.18531056(*) &  -0.21134602(*) &  -0.21942397(*) &  -0.22246746(*) &      -0.22641(23) \\
& RHF      & -15.40096815 & -15.40098612 & -15.40100365 & -15.40100918 & -15.40101005 &  -15.40101079(57) \\
\cline{8-8}
& Total    &              &              &              &              &              &     -15.62740(23) \\
\hline

\hline\hline
\end{tabular}
\end{table*}

%% file: results/fn_bias_tables.tex
\begin{table}[htbp!]
\small
\centering
\setlength{\tabcolsep}{4pt} 
\caption{
This table summarizes the exact/nearly-exact total energies for ccECPs (Rb--In with [[Ar]3$d^{10}]$ core) from CBS extrapolations of CCSDT(Q)/aug-cc-pCV(n)Z calculations, and FN-DMC energies extrapolated to zero time-step. 
Total FN-DMC error ($\epsilon$) (including ccECP locality biases), and percentage error relative to the exact estimation of the correlation energy ($\eta=(\epsilon/|E_{CBS,corr}|)\times100$) are shown. 
FN-DMC calculations used HF nodes in the $D_{2h}$ point group with QZ basis sets, as detailed in Sec.~\ref{subsubsec:dmc}.
The percentage errors are plotted in Fig.~\ref{fig:FN-all} for better illustration of trends.}
\label{tab:FN-Rb-In28}
\begin{tabular}{|c|cc|cc|}
\hline\hline
Atom & exact (Ha)   & SD-DMC (Ha)    & $\epsilon$ (mHa) &  $\eta$(\%) \\
\hline
Rb & -24.10207(14) & -24.09667(15) & 5.4(2) & 2.03(8) \\
Sr28 & -30.52887(16) & -30.52272(34) & 6.2(4) & 2.0(1) \\ 
Y & -38.19409(59) & -38.17932(33) & 14.8(7) & 3.9(2) \\
Zr & -46.77375(52) & -46.75449(69) & 19.3(9) & 4.5(2) \\
Nb & -56.78962(29) & -56.77358(32) & 16.0(4) & 3.44(9) \\
Mo & -67.98559(24) & -67.96706(30) & 18.5(4) & 3.57(7) \\
Tc & -80.31662(26) & -80.28794(48) & 28.7(5) & 4.94(9) \\
Ru & -94.49322(91) & -94.46354(73) & 30(1) & 4.4(2) \\
Rh & -110.10443(85) & -110.06981(47) & 35(1) & 4.7(1) \\
Pd & -127.3828(16) & -127.3396(10) & 43(2) & 4.9(2) \\
Ag & -146.94686(71) & -146.90703(61) & 39.8(9) & 4.5(1) \\
Cd & -167.63155(56) & -167.58205(56) & 49.5(8) & 5.22(8) \\
In28 &-190.2362(36) & -190.18156(68) & 55(4) & 5.4(4) \\
\hline
\end{tabular}
\end{table}
\begin{table}[htbp!]\centering
\setlength{\tabcolsep}{4pt} 
\caption{Exact/nearly exact total energies for ccECPs: Sr with [Kr] core and In--Xe with [[Kr]4$d^{10}]$ core, from CBS extrapolations of FCI or CCSDT(Q)/aug-cc-pV(n)Z calculations. All other methodological details follow Table~\ref{tab:FN-Rb-In28} also plotted in Fig.~\ref{fig:FN-all}.
}

\label{tab:FN-Sr36-Xe}
\begin{tabular}{|c|cc|cc|}
\hline\hline
Atom & exact (Ha)   & SD-DMC (Ha)    & $\epsilon$ (mHa) &  $\eta$ \\
\hline
Sr36 & -0.594578(25) & -0.594228(76) & 0.35(8) & 1.3(3) \\
In46 & -1.89786(20) & -1.897093(97) & 0.8(2) & 1.6(5) \\
Sn & -3.343938(42) & -3.34234(14) & 1.6(1) & 2.3(2) \\
Sb & -5.39066(20) & -5.388084(78) & 2.6(2) & 2.8(2) \\
Te & -8.14155(43) & -8.13662(13) & 5.3(4) & 3.9(3) \\
I & -11.39668(99) & -11.391208(65) & 5(1) & 3.0(5) \\
Xe & -15.62740(23) & -15.62026(30) & 7.1(4) & 3.2(2) \\
\hline
\hline
\end{tabular}
\end{table}

%% file: results/kinetic_tables.tex
\begin{table*}[htbp!]
\setlength{\tabcolsep}{4pt} 
\centering
\small
\caption{Atomic Kinetic Energies (Ha) for fifth-row elements (Rb--In) with ccECPs [[Ar]3$d^{10}]$ [aug-cc-pCVnZ Basis Set].
Values with (*) were not feasible to calculate and represent estimates from the calculated data as described in the text.
}
\label{tab:kin_Rb-Mo_ccECPs}
\begin{tabular}{|c|l|llll|l|}
\hline\hline
Atom & Method &          $E_{\text{DZ}}^{\text{kin}}$  &          $E_{\text{TZ}}^{\text{kin}}$ &          $E_{\text{QZ}}^{\text{kin}}$ &          $E_{\text{5Z}}^{\text{kin}}$ &             $E_{\text{CBS}}^{\text{kin}}$ \\
\hline
\multirow{2}{*}{\bf{Rb}}
& ROHF & 7.76204092 & 7.76212048 & 7.76219774 & 7.76222502 &             \\
\cline{7-7}
& CISD & 7.84399114 & 7.92126308 & 7.94415239 & 7.94997286 &  7.9558(29) \\
\hline
\multirow{2}{*}{\bf{Sr}}
& RHF & 9.29447686 & 9.29444848 & 9.29448177 & 9.29450736 &  \\
\cline{7-7}
& CISD & 9.47565511 & 9.50391740 & 9.50851577 & 9.51515701 & 9.51681(41) \\
\hline
\multirow{1}{*}{\bf{Y}}
& ROHF & 12.14382813 & 12.14399966 & 12.14391178 & 12.14389221 &  \\
\cline{7-7}
& CISD & 12.31093706 & 12.39387205 & 12.41633857 & 12.42358605 & 12.4308(36)  \\
\hline
\multirow{2}{*}{\bf{Zr}}
& ROHF & 15.11590542 & 15.11597041 & 15.11590976 & 15.11588221 &  \\
\cline{7-7}
& CISD & 15.31359675 & 15.40631892 & 15.42993214 & 15.43957285 & 15.4492(48) \\
\hline
\multirow{3}{*}{\bf{Nb}}
& ROHF & 19.61639336 & 19.61709419 & 19.61716650 & 19.61714229 &  \\
\cline{7-7}
& CISD & 19.82140400 & 19.94049692 & 19.97219882 & 19.98524687 & 19.9983(65)\\
\hline
\multirow{2}{*}{\bf{Mo}}
& ROHF & 23.88350226 & 23.88346492 & 23.88362478 & 23.88339017 &  \\
\cline{7-7}
& CISD & 24.10056987 & 24.22936091 & 24.26750511 & 24.28386580 & 24.3002(82) \\
\hline
\multirow{3}{*}{\bf{Tc}}
& ROHF & 28.03180660 & 28.03198977 & 28.03197302 & 28.03186577 &              \\
\cline{7-7}
& CISD & 28.22949762 & 28.34965080 & 28.38490494 & 28.40442249 &  28.4239(98) \\
\hline
\multirow{3}{*}{\bf{Ru}}
& ROHF & 34.69716930 & 34.69592898 & 34.69541085 & 34.69522581 &  \\
\cline{7-7}
& CISD & 34.89571035 & 35.06709613 & 35.11386296 & 35.13855936 & 35.163(12) \\
\hline
\multirow{3}{*}{\bf{Rh}}
& ROHF & 41.12207061 & 41.12728723 & 41.12787718 & 41.12784662 &             \\
\cline{7-7}
& CISD & 41.32208202 & 41.52814923 & 41.58574834 & 41.61275484 &  41.640(14) \\
\hline
\multirow{3}{*}{\bf{Pd}}
& RHF & 48.34774340 & 48.34793599 & 48.34728902 & 48.34932191 &  \\
\cline{7-7}
& CISD & 48.62489514 & 48.84897627 & 48.93567615 & 48.99187017 & 49.048(28) \\
\hline
\multirow{3}{*}{\bf{Ag}}
& ROHF & 56.61884627 & 56.61908897 & 56.61895873 & 56.61978533 &             \\
\cline{7-7}
& CISD & 56.82414694 & 57.07166189 & 57.14588159 & 57.18222712 &  57.219(18) \\
\hline
\multirow{3}{*}{\bf{Cd}}
& ROHF & 63.95645185 & 63.95636405 & 63.95644751 & 63.95664326 &             \\
\cline{7-7}
& CISD & 64.14819383 & 64.37268462 & 64.45436025 & 64.50012443 &  64.546(23) \\
\hline
\multirow{3}{*}{\bf{In}}
& ROHF & 73.56942916 & 73.56984588 & 73.56937278 & 73.56976475 &             \\
\cline{7-7}
& CISD & 73.87673197 & 74.16007583 & 74.20123511 & 74.27660042 &  74.352(38) \\
\hline

\hline\hline

\end{tabular}
\end{table*}
\begin{table*}[htbp!]
\setlength{\tabcolsep}{4pt} 
\centering
\small
\caption{Atomic kinetic energies (in Ha) for 5$s$ (Sr with [Kr] core) and 5$p$ (In--Xe with [[Kr]4$d^{10}]$ core) main group elements calculated using ccECPs and aug-cc-pVnZ basis sets. CBS extrapolation employed TZ--6Z series. The remaining conventions are consistent with those in Table \ref{tab:kin_Rb-Mo_ccECPs}.
}
\label{tab:kin_Sr36-Xe_ccECPs}
\begin{tabular}{|l|l|lllll|l|}
\hline\hline
Atom & Method &    $E_{\text{DZ}}^{\text{kin}}$ &      $E_{\text{TZ}}^{\text{kin}}$ &          $E_{\text{QZ}}^{\text{kin}}$ &          $E_{\text{5Z}}^{\text{kin}}$ &          $E_{\text{6Z}}^{\text{kin}}$ &             $E_{\text{CBS}}^{\text{kin}}$ \\
\hline
\multirow{1}{*}{\bf{Sr}}
& ROHF & 0.15694930 & 0.15696631 & 0.15696640 & 0.15696170 & 0.15696453 &  \\
\cline{8-8}
& FCI & 0.17089510 & 0.17391444 & 0.17459015 & 0.17479485 & 0.17467204 & 0.174549(61) \\
\hline
\multirow{2}{*}{\bf{In}}
& ROHF & 0.68435079 & 0.68435542 & 0.68435222 & 0.68436798 & 0.68436165 &              \\
& CISD & 0.70712321 & 0.70825426 & 0.71614225 & 0.71701303 & 0.71726755 &              \\
\cline{8-8}
& FCI  & 0.70884433 & 0.71066666 & 0.71870873 & 0.71966045 & 0.71993417 &  0.72021(14) \\
\hline
\multirow{1}{*}{\bf{Sn}}
& ROHF & 1.15296566 & 1.15306108 & 1.15314026 & 1.15318501 & 1.15312311 &  \\
& CISD & 1.18340987 & 1.19194378 & 1.19628759 & 1.19763665 & 1.19786707 &  \\
\cline{8-8}
& FCI & 1.18665194 & 1.19808956 & 1.20300002 & 1.20447878 & 1.20473807 & 1.20500(13) \\
\hline
\multirow{2}{*}{\bf{Sb}}
& ROHF & 1.76925483 & 1.76936118 & 1.76931047 & 1.76933101 & 1.76927902 &              \\
& CISD & 1.80110385 & 1.81752391 & 1.82364247 & 1.82531592 & 1.82614441 &              \\
\cline{8-8}
& FCI  & 1.80411924 & 1.82663442 & 1.83457312 & 1.83625660(*) & 1.83709006(*) &  1.83792(42) \\
\hline
\multirow{3}{*}{\bf{Te}}
& ROHF & 2.57155426 & 2.57148037 & 2.57147050 & 2.57148720 & 2.57146504 &  \\
& CISD & 2.59525291 & 2.62761479 & 2.64549060 & 2.64945670 & 2.65126112 &  \\
\cline{8-8}
& FCI & 2.59898425 & 2.63996388 & 2.65792370(*) & 2.66190844(*) & 2.66372134(*) & 2.66553(91) \\
\hline
\multirow{3}{*}{\bf{I}}
& ROHF & 3.54929571 & 3.54944174 & 3.54931902 & 3.54941068 & 3.54940790 &             \\
& CISD & 3.57453167 & 3.61307510 & 3.66048321 & 3.66766215 & 3.67037605 &             \\
\cline{8-8}
& FCI  & 3.57849276 & 3.62864329 & 3.67625567(*) & 3.68346555(*) & 3.68619114(*) &  3.6889(14) \\
\hline
\multirow{3}{*}{\bf{Xe}}
& ROHF & 4.49822985 & 4.49359089 & 4.49684704 & 4.49675022 & 4.49687318 &             \\
& CISD & 4.53656729 & 4.59836170 & 4.63225045 & 4.64182050 & 4.64554703 &             \\
\cline{8-8}
& FCI  & 4.54273614 & 4.60461458(*) & 4.63854941(*) & 4.64813247(*) & 4.65186407(*) &  4.6556(19) \\
\hline\hline
\end{tabular}
\end{table*}

\begin{table}[!htbp]
\centering
\caption{ccECP atomic kinetic energies and ratio of kinetic to total (\%) energies for Rb--In with [Ar]3$d^{10}$ cores.
}
\label{tab:KE-Rb-In28}
\begin{tabular}{|c|cc|}
\hline\hline
Atom & $E_{\text{CBS}}^{\text{kin}}$ (Ha)  &   $E_{\text{CBS}}^{\text{kin}}/|E_{total}|$(\%) \\
\hline
Rb  &  7.9558(29) &  33.01(1)  \\
Sr  &  9.51681(41) &  31.173(1) \\
Y   & 12.4308(36) & 32.546(9) \\
Zr  &   15.4492(48) &  33.03(1) \\
Nb  &   19.9983(65) &  35.21(1) \\
Mo  &   24.3002(82) &  35.74(1) \\
Tc  &  28.4239(98) &  35.39(1)\\
Ru  &    35.163(12) &  37.21(1) \\
Rh  &  41.640(14) &  37.82(1) \\
Pd  &  49.048(28) &  38.50(2) \\
Ag  &  57.219(18) &  38.94(1) \\
Cd  &  64.546(23)  &  38.50(1)  \\
In  &  74.352(38)  &  39.08(2) \\
\hline
\hline
\end{tabular}
\end{table}
\begin{table}[!htbp]
\centering
\caption{ccECPs atomic kinetic energies and ratios of kinetic to total (\%) energies for Sr with [Kr] core and In--Xe with [Kr]3$d^{10}]$ cores.
}
\label{tab:KE-Sr-Xe}
\begin{tabular}{|c|cc|}
\hline\hline
Atom & $E_{\text{CBS}}^{\text{kin}}$ (Ha)  &   $E_{\text{CBS}}^{\text{kin}}/|E_{total}|$(\%) \\
\hline
Sr  & 0.174549(61) &  29.36(1) \\
In  & 0.72021(14) &  37.949(8) \\
Sn  & 1.20500(13) &  36.035(4) \\
Sb  & 1.83792(42) &  34.095(8) \\       
Te  &  2.66553(91) &  32.74(1) \\
I   &  3.6889(14) &   32.37(1) \\
Xe  &  4.6556(19) &  29.79(1) \\
\hline
\hline
\end{tabular}
\end{table}

%% file: conclusions.tex
\section{Conclusions}\label{conclusion}

As methods for calculating electronic structure become more accurate, there is a growing need for effective valence-only Hamiltonians that are tested, benchmarked, and referenced for basic properties like total energies. We note that total energies for all-electron systems have been studied in detail using high-accuracy methods such as coupled cluster, multi-determinant DMC and even FCI; relevant benchmarks are provided in works like \cite{scemama-3d,QMC_AE,troger-C,helium}.
In this work, we extend the benchmarking of ccECPs by providing accurate data for the next series of elements (Rb$-$Xe) using the accompanying basis sets. We also include accurate kinetic energy estimations and, to aid with comparison, the variance of the local energies (see the supplementary material tables S.7-S.8). We also emphasize the utility of ccECPs which provide multiple core-size options to best balance accuracy and computational cost for applications in large systems.

We analyze data trends, particularly from the perspective of QMC methods that are applicable to larger systems. We study the present trends in fixed-node errors for transition metals, contrasting $3d$ versus $4d$ elements. The results confirm that increases in fixed-node error correlate with higher electron densities and with the addition of the first orbitals of a new angular momentum. By the same reasoning, this trend also relates to an increase in kinetic energy which originates from a corresponding increase in wavefunction curvature and more complex nodal shapes in these regions.
Furthermore, we analyze the kinetic-to-total energy ratio to assess the smoothing of valence states and its trend across the heavier elements. The increasing utilization of ccECPs in large systems applications (e.g., molecular chains, solids, polymers) using methods like \textit{ab initio}, QMC, and neural-network based approaches provide robust validation of their suitability for such tasks\cite{ccECP-use-1,ccECP-use-2,ccECP-use-3,ccECP-use-4,ccECP-use-5,ccECP-use-6}. The transparent residual errors in ccECPs offer practical guidance for high-accuracy studies. We note that these residual biases could be further reduced in the future through more refined constructions, as interest and as need dictates.